\documentclass[final,authoryear,5p,times,twocolumn]{elsarticle}

\usepackage{graphicx}

\usepackage{amssymb}
\usepackage{amsmath}

\usepackage[nodots]{numcompress}

\usepackage{lineno}

\usepackage{color}
\usepackage{url}
\usepackage{verbatim}
\usepackage{bm}
\usepackage{setspace}

\usepackage[final] {changes}

\journal{Journal of Theoretical Biology}

\begin{document}

\begin{frontmatter}



\title{Energy cost and optimisation in breath-hold diving}


\author{M. Trassinelli}
\address{Institut des NanoSciences de Paris,  CNRS-UMR 7588, Sorbonne Universit\'es, UPMC Univ Paris 06, 75005, Paris, France}
\ead{trassinelli@insp.jussieu.fr}

\begin{abstract}

We present a new model for calculating locomotion costs in breath-hold divers.
Starting from basic mechanics principles, we calculate the work that the diver must provide through propulsion to counterbalance the action of drag, the buoyant force and weight during immersion.
\deleted{The basal metabolic rate and the efficiency to transform chemical energy in propulsion are also considered for the calculation of the total energy cost of a dive.} 
Compared to those in previous studies, the model presented here accurately analyses breath-hold divers which alternate active swimming with prolonged glides during the dive (as is the case in mammals). 
\added{The energy cost of the dive is strongly dependent on these prolonged gliding phases. 
Here we investigate the length and impacts on energy cost of these glides with respect to the diver characteristics, and compare them with those observed in different breath-hold diving species.} 
\deleted{The extension of the gliding phases and their impact on the total energy cost of the dive are investigated and compared to different breath-hold divers observations.}\added{Taking into account the basal metabolic rate and chemical energy to propulsion transformation efficiency, we calculate optimal swim velocity and the corresponding total energy cost (including metabolic rate) and compare them with observations.}
\deleted{For different dive conditions, the optimal swim velocity and the corresponding energy cost (that includes the metabolic rate) are also calculated and compared with experimental data of breath-hold diving animal.}
\added{Energy cost is minimised when the diver passes through neutral buoyancy conditions during the dive. This generally implies the presence of prolonged gliding phases in both ascent and descent, where the buoyancy (varying with depth) is best used against the drag, reducing energy cost. This is in agreement with past results \citep{Miller2012,Sato2013} 
where, when the buoyant force is considered constant during the dive, the energy cost was minimised for neutral buoyancy. In particular, our model confirms the good physical adaption of dolphins for diving, compared to other breath-hold diving species which are mostly positively buoyant (penguins for example). The presence of prolonged glides implies}
a non-trivial dependency of optimal speed on maximal depth of the dive\deleted{is found}. 
\added{This extends previous findings \citep{Sato2010,Watanabe2011} which found no dependency of optimal speed on dive depth for particular conditions.}
The energy cost of the dive \replaced{can be further diminished by reducing the volume of gas-filled body parts}{results minimised} in divers close to neutral buoyancy. \deleted{and with a reduced volume of body parts filled with gas}
This provides a possible additional explanation for the observed exhalation of air before diving in phocid seals \added{to minimise dive energy cost. Until now the only explanation for this phenomenon has been a reduction in the risk of decompression sickness}.

\end{abstract}

\begin{keyword}
swimming \sep cost of transport \sep response to diving \sep buoyancy \sep dolphin
\end{keyword}

\end{frontmatter}


\section{Introduction}
During their dives, breath-hold diving animals minimise energetic cost to gain time foraging  as oxygen stored in their body is limited.  
Besides plastic physiological adaptations to diving, like bradycardia, reduction and redistribution of the blood flow, etc., \citep{Butler1997,Kooyman1985,Kooyman1998,Butler2004}, dive energy cost can be lowered by reducing dive duration and/or the mechanical work necessary for propulsion. 
Energy cost related to the basal metabolic rate is proportional to dive duration and inversely proportional to swimming velocity. On the other side, energy spent for propulsion depends on the drag force during the dive, which increases with the square of velocity. Besides swimming optimisation and hydrodynamics, thrust work is efficiently reduced by slowing down swim speed. Optimal dive velocity is a compromise, taking into account these two elements and is specific to the diver's body characteristics and maximal dive depth. An additional energy cost reduction is obtained, particularly in mammals, by alternating active swimming with prolonged glides using buoyant force and weight to their advantage but also by varying their stroke frequency and/or by using stroke-and-glide swimming \citep{Crocker1997,Webb1998,Skrovan1999,Williams2000,Davis2001,Nowacek2001,Biuw2003,Sato2003,Miller2004,Watanabe2006,Aoki2011,Maresh2014}.

In recent years many models have been developed to study energy economy in breath-hold divers, mainly birds and mammals \citep{Wilson1992,Skrovan1999,Hansen2004,Sato2010,Aoki2011,Watanabe2011,Miller2012}.
). In these studies, energy cost is estimated from first principles from the work of the mechanical forces acting on the diver and the diver's metabolism. 
A complete force and energy cost analysis has been made for penguins, which are usually positively buoyant and glide only during the ascending phase,  \citep{Hansen2004,Sato2010,Watanabe2011,Miller2012}.
In particular \citet{Sato2010,Watanabe2011} found the analytical expression of optimal velocity for minimising energetic cost of the dive \added{for this particular case}.
\deleted{A numerical approach to the problem is also presented in Miller et al. (2012).}

\replaced{In the case of mammals, the problem is more complex due to}{No similar analysis is available for mammals, where} their alternation of prolonged glides and active swimming phases  -- crucial for energy balance.
\added{For this case, an exhaustive analysis is still missing in the literature.
An investigation similar to \citet{Sato2010} for mammals was developed by \citet{Miller2012}), though the dependency of buoyancy on depth was not taken into account.}
 \citet{Skrovan1999} produced another analysis where only the average value of the buoyant and thrust forces was considered. 
A numerical study was also developed by \citet{Davis2007} for elephant \replaced{seals}{lions}, but in this case it was for a related investigation where the benefit of transit dives with gliding descent is compared to purely horizontal displacement.

Here we present a new model for breath-hold \replaced{diving}{divers} for a general case which can include the presence of prolonged gliding phases in both ascent and descent. 
\deleted{Starting from the analysis of the involved mechanical forces}
\added{We begin with analysis of the mechanical forces involved and their exact dependency on depth. We calculate length of the gliding phases  and take it into account to evaluate dive locomotion cost.
Moreover, by including}
\deleted{, their dependency on depth and }
some basic consideration of the animal physiology, we also predict the energy cost of a typical dive.
\deleted{The extension of the gliding regions is calculated and is included in the locomotion cost of the dive.}
We study the dependency of locomotion and energy cost on the different parameters (dive velocity, diver buoyancy, mass, etc.) and their optimisation. We then compare this with observations of real dives in different breath-hold diving animals, in particular the bottlenose dolphin, for which several \added{vertical-dive} measurement from trained individuals are available

In the following section of this paper we present our model to calculate mechanical work and total energy cost required for a typical dive as a function of the diver and the dive characteristics, including animal metabolism. In the third section we discuss our results and how they compare with observations of real dives. Section four is our conclusion.

\section{Model description}

\subsection{Basic assumptions}
\begin{figure}
\begin{center}
\includegraphics[width=0.4\columnwidth]{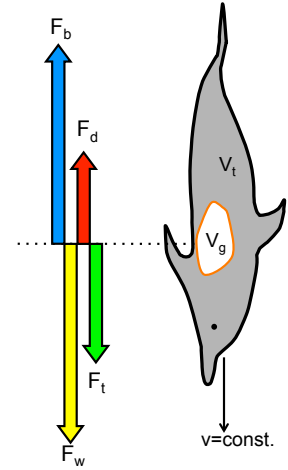} 
\caption{Scheme of the forces acting on a breath-hold diver (a dolphin in the sea in this example) moving with a velocity $v$ in a liquid. Forces are: thrust $F_t$, drag $F_d$, buoyant force $F_b$ and weight $F_w$.
$V_g$ indicates the volume of the gas-filled parts of the diver body (mainly the lung, oral cavities, etc.) compressible with pressure changes.
$V_t$ indicates the volume of the liquid and solid parts of the diver body (body tissues), which do not change significantly in volume when the pressure increases.}
\label{fig:dolphin}
\end{center}
\end{figure}

\added{We calculate dive energy cost from first principles from the work of the mechanical forces acting on the diver and their dependency on depth, as did 
 \citet{Wilson1992,Skrovan1999,Hansen2004,Sato2010,Aoki2011,Watanabe2011,Miller2012}.}
During a dive in a fluid (sea water in our case), the animal is subject to four forces: drag, weight, the buoyant force and thrust (see scheme in Figure~\ref{fig:dolphin}).
Drag is caused by friction and the pressure gradient differential. It is strongly dependent on the body velocity $v=|\boldsymbol{v}|$ and always opposes it ($\boldsymbol{F_d} = - F_d(v) \boldsymbol{\hat v} $ with $\boldsymbol{\hat v} = \boldsymbol{v}/v$). 

Weight is proportional to body mass $m$ and the acceleration of free fall $g$ ($\boldsymbol{F_w} = m \boldsymbol{g}$). 
The buoyant force is its counterpart and is proportional to body volume $V$ and the fluid density $\rho$  ($\boldsymbol{F_b} = -\rho V \boldsymbol{g}$). 
The buoyant force changes with depth due to the presence of gas-filled parts of the diver's body (lungs, oral cavities, etc.) whose volume $V_g$ varies with the pressure $P$ ($P V_g = $ const.).
To take into account the effect of $V_g$ variations, it is better to rewrite $F_b$ as a function of the depth $d$ as:
\begin{equation}
F_b = \rho(V_t + V_g) g =  m g \left(R_t + \frac{R_g}{1+\frac{\rho g d
    }{ P_0}}\right),
\label{eq:Fb}
\end{equation}
where $V_t$ represents the 
volume of the tissue of the diver's body comprising liquids and solids which is considered constant at any depth, due to its small variation with pressure\added{, as well as the water density $\rho$}.
We also introduce the ratios $R_t=\rho V_t / m$ and $R_g=\rho V^0_g / m$, where $V^0_g$  represents the gas-filled body parts volume at the surface. 
$R_t$ can also be written with respect to fluid density $\rho$ and average body tissue density $\rho_t$: $R_t=\rho / \rho_t$.
The use of the ratios $R_t$ and $R_g$ instead of the volumes $V_t$ and $V_g$ has the advantage of allowing us to investigate the dependency of locomotion cost and other pertinent quantities on the different parameters independently (or almost) from the specific mass of the diver.

\begin{table}[h!]
\centering
\begin{tabular}{l l l} 
\hline
Symbol & Meaning & Units \\
\hline
$m$	& mass of the diver & kg \\
$D$ & maximum depth of the considered & m\\
& dive \\
$T$ & total diving time & s\\
$\boldsymbol{v}$ & swim velocity & m s$^{-1}$\\
$\boldsymbol{F_d}$ & drag force & N\\
$\boldsymbol{F_w}$ & weight & N\\
$\boldsymbol{F_b}$ & buoyant force & N\\
$\boldsymbol{F_t}$ & thrust & N\\
$C$ & drag constant & Kg m$^{-1}$\\
$\lambda$ & ratio between active to \\
& passive swimming drag\\
$\boldsymbol{g}$ & free fall acceleration & m s$^{-2}$\\
$P_0$ & pressure at the surface & Pa \\
$\rho$ & liquid density (here the sea density  & Kg m$^{-3}$ \\
& assumed to be 1027~Kg m$^{-3}$) \\
$\rho_t$ & diver body tissue average density & Kg m$^{-3}$\\
$V_g$ & volume of the body parts & m$^{3}$\\
 & filled with gas (compressible) \\
$V^0_g$ & volume of the gas-filled  & m$^{3}$ \\ 
& body parts at the surface \\
$R_g$ & $=\rho V_g / m$, ratio between the \\
& gas-filled body volume and the mass\\
$V_t$ & volume of the body tissue & m$^{3}$\\
&  (assumed incompressible) \\
$R_t$ & $=\rho V_t / m=\rho/\rho_t$, ratio between the \\
& tissue body volume and the mass\\
$d_{D}$ & depth where $F_b + F_d = F_w$ & m \\
& (descent critical depth)\\
$d_{A}$ & depth where $F_w + F_d = F_b$ & m\\
& (ascent critical depth)\\
$W$ & work of the thrust for the dive & J \\
& (locomotion cost of the dive)\\
$\varepsilon_p$ & efficiency to transform muscular\\
&  movement into propulsion \\
$\varepsilon_m$ & efficiency to transform chemical\\
& energy  into muscular movement \\
$\varepsilon$ & $=\varepsilon_p \varepsilon_m $, efficiency to transform\\
& chemical energy into thrust \\
$E$ & $=E_m + E_t$, total chemical energy  & J\\
& required for the dive \\
$E_t$ & chemical energy required & J\\
& for the thrust \\
$E_m$ & chemical energy required for  & J\\
& the basal metabolism\\
$B$ & basal metabolic rate & W \\
COT & $=E/(m D)$, cost of transport & J m$^{-1}$ Kg$^{-1}$\\
\hline
\end{tabular} 
\caption{List of symbols with corresponding units. Bold symbols indicates vectorial quantities. The corresponding symbols not in bold indicate their absolute values. All quantities are in the units of the international system of units (SI). Quantities without units are dimensionless.}
\label{table:symbols}
\end{table}

Locomotion cost required to descend to a maximal depth $D$ and return to the surface can be quantified by the calculation of the work $W$ of the thrust force produced by the diver along to the dive path. 
In order to reduce $W$, breath-holding animals (and also human freediver athletes) use the buoyant force and weight in their favour by gliding for as long as possible. 
In particular, some mammals such as dolphins, seals, etc. glide for large parts of both descending and ascending dive phases. 
For these mammals, a typical dive is composed by a first step where the diver makes an effort to descend from the surface.  
Then it actively swims down to a depth $d_{D}$ (descent critical depth), where the buoyancy is small enough to allowing it to glide to the maximum depth $D$. 
Following this, the diver makes another effort to ascend from the maximum depth to a second depth $d_{A}$ (ascent critical depth) from which it glides to the surface under the action of the buoyant force.
The mechanical work for the round-trip can be formally written as
\begin{equation}
W(D) = \int_0^{d_{D}} \boldsymbol{F_t}(d) \cdot \boldsymbol{d \ell}
+ \int_D^{d_{A}} \boldsymbol{F_t}(d) \cdot \boldsymbol{d \ell},
\label{eq:Wbasic}
\end{equation}
where $\boldsymbol{d \ell}$ is the infinitesimal of the dive path.

\subsection{Additional assumptions and approximations}
Several simplifications and assumptions are necessary to study this formula and compare its predictions with observations. These assumptions are discussed in the following sections.

\subsubsection{Trajectory and swim velocity during the dive}
To calculate dive locomotion cost, we consider purely vertical dives, without horizontal displacement and without considering possible foraging at the bottom or intermediate depths. 
We assume in addition that the diver travels at a constant velocity for the descent and the ascent, which is partially justified by the observation of velocity profiles of many breath-hold mammals' dives \citep{Skrovan1999,Williams2000,Biuw2003}.  
\added{We must note that dives in natural conditions are rarely exclusively vertical and they may also have a non-constant swim velocity profile.
But for a given maximal depth, a vertical trajectory minimises the locomotion cost. Analysis of vertical dives then provides maximal or minimal values of the involved quantities (total locomotion cost, gliding phase extension, etc.) for more general dive conditions.
Moreover, understanding this simple case provides the basis for a possible generalisation to dives with an arbitrary dive pitch angle and velocity profile.}

\replaced{The assumption of a}{A} constant velocity implies no acceleration, i.e. the resulting total force is equal to zero. 
Given that, the thrust  $F_t$ must balance the other forces with
\begin{equation}
\boldsymbol{F_t}(d) = -(\boldsymbol{F_b}(d)+\boldsymbol{F_w}+\boldsymbol{F_d}), \label{eq:force}
\end{equation} 
where dependency on depth comes only from buoyancy.

IIn the gliding phases of the dive, velocity tends to increase due the effect of buoyant force variation with depth. It could be kept constant with an intentional increasing of drag without any additional significant effort. A deviation from the assumption of constant velocity during the glides does not, however, affect the consideration of locomotion cost. On the contrary, it can influence the total metabolic energy cost evaluation for the dive, which could be overestimated.

\subsubsection{Swim technique and accelerations}
Besides active swimming with a continuous series of strokes or prolonged glides, a third swim mode, called stroke-and-glide (a short acceleration phase produced by a single or a small series of strokes followed by a small glide), is commonly observed in breath-hold divers   \citep{Webb1998,Skrovan1999,Nowacek2001,Sato2003,Miller2004,Watanuki2005,Watanabe2006,Aoki2011,Sato2011,Shiomi2012}.
Here we consider only the resulting averaged thrust and the corresponding average velocity neglecting possible short local accelerations present in the stroke-and-glide swim mode.
In the same way,  we do not consider the work required at the beginning of the dive to accelerate to the cruise velocity and, at the maximum depth, to change direction, which is about $3/2\ m v^2$, generally much smaller than the total work.

\subsubsection{Passive and active drag}
For typical swim velocities and body masses of breath-hold divers, turbulent regime must be considered \citep{Gazzola2014}.
Drag is then proportional to the square of the velocity, $F_d(v) = C v^2$. 
$C$ is a constant that includes dependency on diver body characteristics and the Reynolds number at this velocity regime. 
Possible additional deviations of drag dependency on the square of velocity  are not considered.
For a given cruise velocity, the drag force can be different if the diver is actively swimming or is gliding. Generally this difference is represented by the parameter $\lambda$ equal to the ratio of active to passive drag.
In many investigations, a value $\lambda > 3$ is considered \citep{Lighthill1971,Webb1975,Fish1998,Skrovan1999,Anderson2001,Fish2014}.
However, studies that include the metabolic rate for locomotion and thermoregulation at the same time indicate that the drag during the active swimming phases should be smaller than during the gliding phases \citep{Hind1997} with $0.2<\lambda<1$ (depending on the diver species).
Experiments with a robotic fish model measure also produce a value of $\lambda < 1$ \citep{Barrett1999}, which is confirmed by recent numerical investigations on hydrodynamic \citep{Borazjani2008}.
In our model, we consider the conservative (and simple) assumption $\lambda = 1$, as in to \citet{Miller2012}.
A deviation from this assumption will be discussed in the following sections.

Unlike \citet{Hansen2004}, which investigated Adelie penguin dives, we do not take into account a possible dependency on depth $d$ of the drag constant $C$ (passive to active). 
This dependency is particularly important for divers with a large portion of their total volume filled with air or other gases, as in the case of penguins. Mammals, the particular focus of this article, have a much smaller relative gas-filled volume, and so are much less subject to variations of $C$ with respect to $d$.

\subsection{Locomotion cost calculation}
From Eqs.~(\ref{eq:Fb}--\ref{eq:force}), and the considerations above, the propulsion work can be written as
\begin{multline}
W(D) 
= [F_d -m g (1-R_t)]d_{D} + \\
+  \frac { m R_g P_0} \rho \log \left(1 + \frac{\rho g\
    d_{D}}{P_0} \right) + \\
+ [F_d +m g (1-R_t)](D - d_{A}) + \\
+\frac { m R_g P_0} \rho \log \left(\frac{1 + \frac{\rho g\ d_{A}}{P_0} }{1 + \frac{\rho g D}{P_0}}\right),
\label{eq:W_gen}
\end{multline}
where $m$ is the diver mass and $R_g$ and $R_t$ are the gas-filled diver body volume and tissue volume over mass ratios, respectively.
$P_0$ is the pressure at the surface, $\rho$ is the water density, $g$ is the free fall acceleration constant and  $d_D$ and $d_A$ are the critical depths where the diver starts to glide in the descent and the ascent, respectively.

Depending on the maximum depth of the dive $D$ and the diver characteristics, the critical depths exist if
\begin{eqnarray}
R_t + \frac{R_g}{1 + \frac{\rho g D}{P_0}} +\frac{ F_d}{m g} < 1 < R_t +
R_g+\frac{ F_d}{m g}  \quad \mbox{for } d_{D},  \label{eq:d-cond1} \\
R_t + \frac{R_g}{1 + \frac{\rho g D}{P_0}} -\frac{ F_d}{m g} < 1 < R_t +
R_g-\frac{ F_d}{m g}   \quad \mbox{for } d_{A}. \label{eq:d-cond2}
\end{eqnarray}
If these conditions are satisfied, the descent and ascent critical depths $d_{D}$ and $d_{A}$ are
\begin{equation}
d_{D,A} = \frac {P_0} {\rho g} \cfrac{ R_g + R_t - 1 \pm \frac{F_d}{mg}} {1 - R_t \mp \frac{F_d}{mg}}.
\label{eq:deq}
\end{equation}

By observing whether or not the critical depths exist, we can distinguish different cases of dive:
\begin{description}
\item[Case A] Both $d_{D}$ and $d_{A}$ exist (the diver partially glides in the descent and ascent).
\item[Case B] Only $d_{D}$ exists (the diver never glides in the ascent).
\item[Case C] Only $d_{A}$ exists (the diver never glides in the descent).
\item[Case D] No critical depth exists and $F_w>F_b+F_d$ for any
  considered depths (the diver is always negatively buoyant, and glides for the entire descent).
\item[Case E] No critical depths exists and $F_b>F_w+F_d$ for any considered depth (the diver is always positively buoyant, and glides for the entire ascent).
\item[Case F] No critical depths exist and
  $F_f>\max(F_b-F_w, F_w-F_b)$  for any considered depth (no gliding phase is present, the drag is dominant). 
\end{description}

In case A, where both critical depths exist, from the combination of Eq.~(\ref{eq:W_gen}) with Eq.~(\ref{eq:deq}) we can obtain the explicit formula of $W$.  
Similar expressions of $W$ for cases B and C can be obtained by substituting one of the critical depths with one of the extremes of the $[0,D]$ interval: $d_{A}=0$ for case B and $d_{D}=D$ for case C. 

For cases D and E, the diver glides for the entire descent or ascent, respectively, and it never glides in the opposite phase ($d_D$ and $d_A$ do not exist) leading to the simple formulae 
\begin{equation}
W^{D,E}(D) = F_d D \pm m g (1-R_t) D \mp \frac {m P_0 R_g}{\rho} \log \left( 1 + \frac{\rho g D}{P_0} \right).
\label{eq:w_caseDE}
\end{equation}

Case F is a peculiar instance where the drag is so strong (or the diver is so close to neutral buoyancy and $R_g \ll 1$) that no gliding phase is present at all.  The work expression for this case is even simpler than cases D and E and it can be formally obtained from $W^D + W^E$ leading to 
\deleted{$W^F(D)=2 F_d D$.}
\added{ 
\begin{equation}
W^F(D)=2 F_d D. \label{eq:w_caseF}
\end{equation}
}
We note that in this case the work for a vertical dive is the same that as for an equivalent horizontal displacement, where only the drag is relevant. Case F is also complementary to case A. The selection of the case (A or F) that must be considered depends on the values of the various parameters It can be demonstrated that if
\begin{equation}
F_d > \frac{\rho g D}{2 P_0} \frac{m g R_g}{1 + \frac{\rho g D}{ P_0}} 
\label{eq:drag_crit}
\end{equation}
the existence of both $d_{D}$ and $d_{A}$ simultaneously in the interval $[0,D]$ is not possible for any value of $R_t$ and so case A is not allowed.

\subsection{Total metabolic energy cost and its optimisation}\label{sec:totalE}
To consider the metabolic cost of the dive, rather than locomotion cost,
the relevant quantity is total energy $E$ which includes basal metabolic rate and the efficiency to transform chemical energy to thrust. $E$ can be broken down into two parts $E=E_m + E_t$.
$E_m = B\ T$ is proportional to basal metabolic rate $B$ and the elapsed time during the dive $T$.  $E_t = W/ \varepsilon$ is proportional to dive mechanical work over the efficiency $\varepsilon$ to transform the chemical energy into forward thrust.  
$\varepsilon=\varepsilon_p  \varepsilon_m$ depends on  the  propulsion efficiency $\varepsilon_p$  to  transform muscular movement into thrust and on metabolic efficiency $\varepsilon_m$ to transform chemical energy into muscular work. 

Following these considerations we can write the expression for total energy $E$ cost for a dive as:
\begin{equation}
E (D,v)=  B\ \frac {2 D} v+ \frac {W(D,v)} \varepsilon,
\label{eq:Etot}
\end{equation}
remembering that we assume a constant swim velocity $v$ during the dive whose resulting duration is $T = 2 D/v$ and where we consider that efficiency $\varepsilon$ does not depend on the swim speed.
In addition, we make explicit the dependency of thrust work $W$ on the velocity. \deleted{ from which $d_{D}$ and $d_{A}$ values depends.}

\section{Results and Discussion}

\subsection{Locomotion work \label{sec:resW}}
Some general considerations can be made on the formulae of $W$ for the different cases.  For any expression of $W$ a linear contribution of $F_d$ is present, as for a horizontal displacement. 
For the simple cases D and E a linear dependency on $m, R_g$ and  $R_t$ also exists.  
For cases A, B and C there is a more complex dependency on these parameters because of the presence of $d_{D,A}$ (Eqs.~(\ref{eq:deq})). 
Variation of $R_g, R_t$ and $F_d/(m g)$ values imply the existence, or not, of the two critical depths and so determines the selection of the case which must be considered for the work calculation. 

Compared to the other parameters, only $F_d$ can be easily varied modulating the swim speed. 
Both $R_g$ and $R_t$ are specific to the diver's body characteristics and have a strong influence on the locomotion cost $W$.
The ratio $R_g=\rho V^0_g/m$ can be modulated by inhaling more or less air before diving and is generally much less than one \citep{Kooyman1998}. 
The ratio $R_t$, equal to the ratio between the sea water and average body tissue densities $\rho/\rho_t$, is generally close to one.  
$R_t$ can be tuned by varying the diver's bodily lipid content, as seasonally happens in elephant seals \citep{Biuw2003,Aoki2011}.
From Eqs.~(\ref{eq:deq}) we see that small variations of $R_t$ can drastically change the value of the critical depths and dive work. 

To compare the model predictions with observations, we consider the case of bottlenose dolphin, for which observations from trained individuals performing purely vertical dives without foraging are available \citep{Williams1999,Skrovan1999}, i.e. the ideal conditions in which to test our model, and where all cases A--F are seen.
To investigate the dependency of the dive work on these important parameters $R_t$ and $R_g$, we consider $W$ for a particular bottlenose dolphin individual with a mass of 177~Kg, lung volume of 8.5~l and where we artificially change the values of $R_t$ and $R_g$ to the values deduced from the observations $R^\mathit{meas.}_t = 0.9809$ and $R^\mathit{meas.}_g = 0.0493$  \citep{Skrovan1999}.
We also have the value of the drag constant value $C=4.15$~kg~m$^{ -1}$ from \citet{Skrovan1999} for the calculation of $F_d$.

\begin{figure}
\begin{center}
\includegraphics[width=\columnwidth]{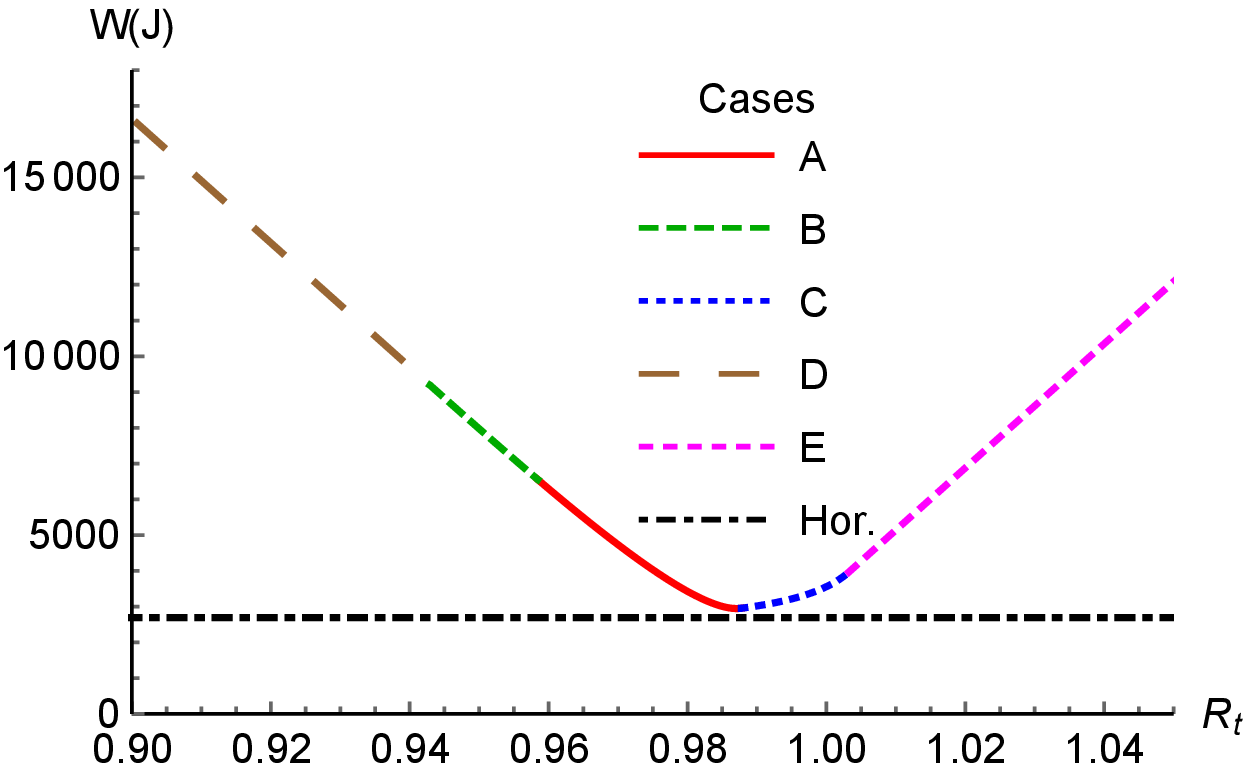}
\includegraphics[width=\columnwidth]{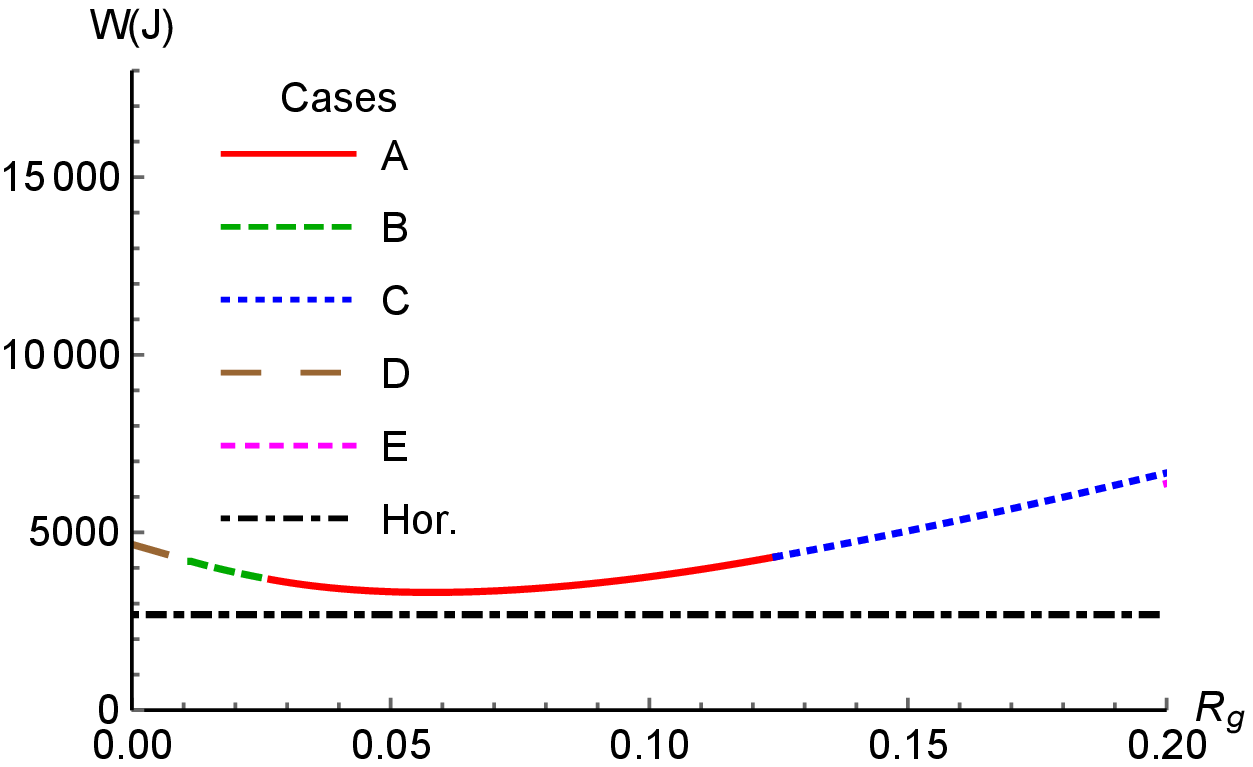} 
\caption{Thrust work for a bottlenose dolphin of mass $m=177$~Kg  and drag constant $C=4.15$~kg m$^{-1}$  diving to a maximum depth of 100~m with a speed of 1.8~m s$^{-1}$. 
Top: Variation of $W$ with respect to $R_t$ keeping $R_g$ at the constant value of 0.0493 ($=R^\mathit{meas.}_g$).
Bottom: Variation of $W$ with respect to $R_g$ keeping $R_t$ at the constant value of 0.9809 ($=R^\mathit{meas.}_t$).
Different colours correspond to different diving cases.} 
\label{fig:work1}
\end{center}
\end{figure}

In Figure~\ref{fig:work1} (top) we consider the locomotion cost for a vertical dive to a maximal depth of 100~m, with a swim velocity of $1.8$~m s$^{-1}$, with $R_g = R^\mathit{meas.}_g$ and where we vary $R_t$.
Different values of $R_t$ correspond to different cases of dive. 
For this particular parameter set, case A is possible (both critical depths can be present for a specific dive).
Similarly, we study the dependency of $W$ on $R_g$ varying its value keeping $R_t = R^\mathit{meas.}_t$ constant (Figure~\ref{fig:work1} bottom).
For variations of $R_g$ we do not take into account its minimal value due to the residual lung volume that, due to alveolar collapse, can be very small at high pressures, \citep{Kooyman1972,Falke1985,Moore2011}.

In both cases, a minimum of  $W$ is present at $R^\mathit{best}_t= 0.9871$  and $R^\mathit{best}_g=0.0578 $ when $R_g$ and $R_t$, respectively, are kept fixed, not far from the real values of the considered dolphin.
Both minima, are included in case A, when both critical depths exist.
More generally, it can be demonstrated that, when case A is possible,  the minimum of $W$ is unique and is always included in case A.
The corresponding critical depths are around the middle-depth $d_{eq D,A} =  D/ 2 \pm \Delta d_{eq}$, where $\Delta d_{eq}$ depends on $m, F_d$ and $R_g$ or $R_t$
\added{and where $\Delta d_{eq} \to 0$ when $F_d \to 0$.
In other words, $W$ is minimised when the diver is neutrally buoyant at the middle-depth of the dive $D/ 2$.}
Out from the minimum, variations of a few percent of the value of $R_t$ lead to changes of W of several orders of magnitude.
When the value of $R_t$ is relatively far from one, i.e. in cases D and E, $W$ increases linearly with it, as expected from Eqs.~(\ref{eq:w_caseDE}).
$W$ is much less sensitive to changes of $R_g$. Variations of several times its value produce moderate changes in the value of $W$.

\begin{figure}
\begin{center}
\includegraphics[width=\columnwidth]{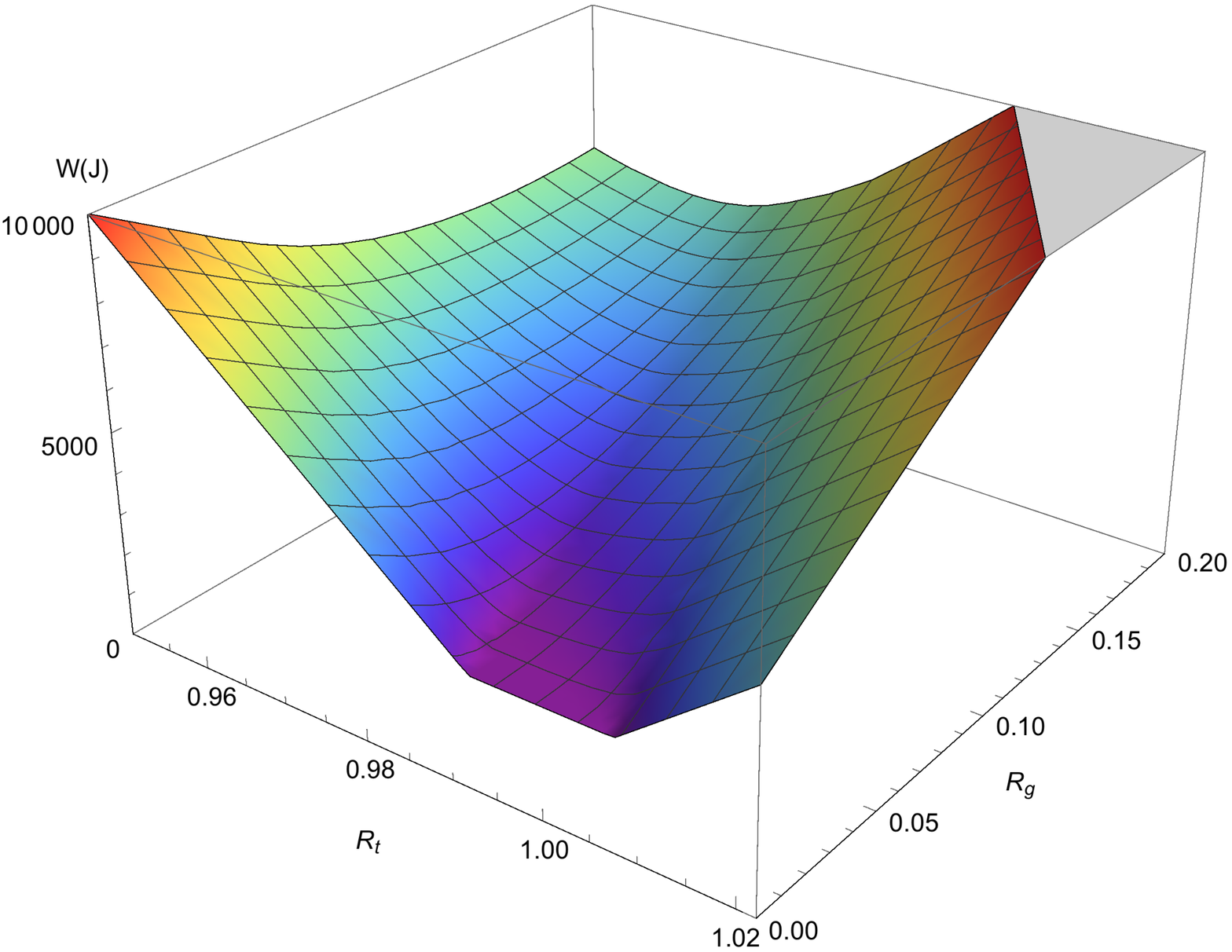} 
\includegraphics[width=\columnwidth]{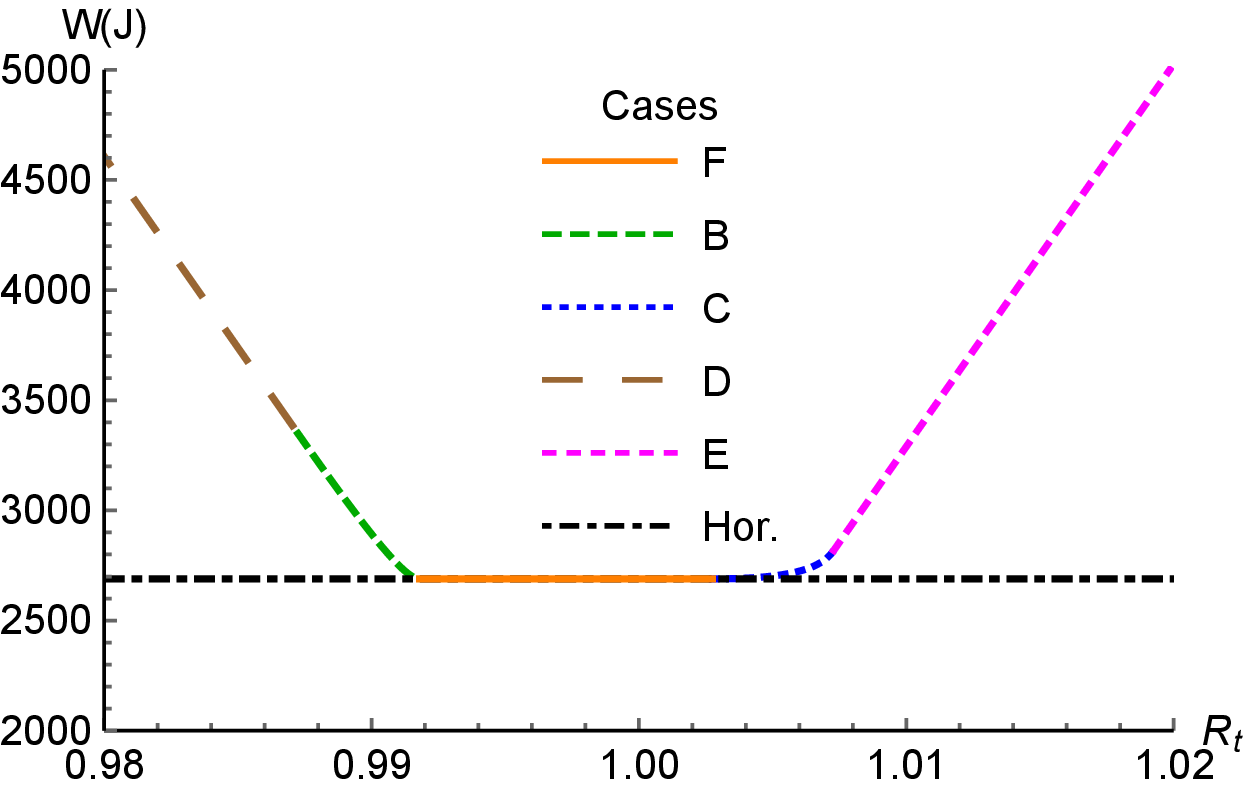} 
\caption{Top: Thrust work as function of $R_t$ and $R_g$ for a bottlenose dolphin of mass $m=177$~Kg and drag constant $C=4.15$~kg m$^{-1}$ diving to a maximum depth of 100~m with a speed of 1.8~m s$^{-1}$. 
Bottom: a constant $R_g$ curve ($R_g =0.005$) close to the minimal $W$ region. \label{fig:work3D}}
\end{center}
\end{figure}

An overall view of the locomotion cost dependency on diver characteristics is presented in Figure~\ref{fig:work3D} (top) where we plot $W$ as a function of both the $R_t$ and $R_g$ parameters.
Following the grid lines corresponding to constant $R_t$ values,  equivalent to the plot in Figure~\ref{fig:work1} (bottom), we can observe that the value $R^\mathit{best}_g \to 0$ when $R_t \to 1$. \deleted{ from values less than unity.}
This is connected to the presence of the global minimal region of $W$ close to the values $R_t=1$ and $R_g=0$.
When $R_g \sim 0$ and $R_t \sim 1$, Eq.~(\ref{eq:drag_crit}) is satisfied and case A is replaced by case F whose range corresponds exactly to the global minimal region of $W$ where $W = 2 D F_d$, and no gliding phases are present in the dive.
A close-up of this global minimal region is presented in Figure~\ref{fig:work3D} (bottom) with the cut curve $W(R_t)$ corresponding to the constant value $R_g =0.005$.
A similar behaviour, with a minimal region corresponding to case F, can also be found for a high drag force (high swim velocity) and/or dives to a limited maximal depth, for which Eq.~(\ref{eq:drag_crit}) can also be satisfied.
In the extreme condition where $R_g=0$, only cases D, E and F are possible (no presence of critical depths), and $W$ is minimal when
\begin{equation}
1- \frac {F_d} {m g} < R_t < 1+ \frac {F_d} {m g} \label{eq:Ri_Rc0}
\end{equation}
is satisfied (case F).

The above results about the minimisation of the thrust work $W$ can also be understood qualitatively with a simple reasoning:
during the dive, thrust has to counterbalance the action of drag, which is a non-conservative force, and weight and its counterpart the buoyant force, which are both conservative forces.
The work spent against the conservative forces in one phase (ascent and/or descent) is given back in the opposite phase of the dive and is used for gliding. 
During the glides, the buoyant force increases or decreases progressively with the change of depth, producing diver acceleration\added{, as observed in penguins \citep{Sato2002},} with a consequent increase in drag. 
It is possible for the diver to maintain constant velocity, but in this case it must increase drag intentionally.
In other words, during the glides, the potential energy related to gravitation is dissipated in the turbulent motion of the water and then lost.
To minimise this, the glide regions should be reduced as much as possible.
This means that  the critical depths should be kept as close as possible to the middle-depth $D/2$
\added{corresponding to condition of neutral buoyancy of the diver around $D/2$,} far from the conditions with a very prolonged glide in one phase of the dive and another short glide or no glide at all in the other phase (like in case D and E).
This is exactly what we found formally from the formulae plots presented above. 
When $R_g$ (or $R_t$) is kept constant, the values $R^\mathit{best}_t$ (or $R^\mathit{best}_g$) that minimises $W$ corresponds to the critical depths $d_{eq D,A} =  D/ 2 \pm \Delta d_{eq}$ \added{and neutral buoyancy of the diver at exactly half the maximal depth of the dive $D/ 2$.
Note that this is in fact a well known practice in freediving (a human sport).}
Where both $R_g$ and $R_t$ are varied, the best values are $R_g \sim 0$ and $R_t \sim 1$, which implies case F with total absence of glides \added{and close to neutral buoyancy}.

The previous consideration implies that when $R_t$ value is sufficiently close to one, a decreasing of thrust work can be obtained by exhaling before diving, to reduce the value of $R_g$.
This operation also sensitively reduces the amount of available stored oxygen, penalising the maximum achievable depth. 
\added{This is not true in the case of phocid seals, where stored oxygen is mainly located in the blood and muscular tissue and where the lung air represents only 5\% of available oxygen reserves \citep{Kooyman1985}.}
Phocid seals \deleted{that} are \added{in fact} observed to exhale before diving \citep{Scholander1940,Kooyman1970}.
This practice is generally interpreted as a reduction in risk of decompression sickness via alveolar collapse \citep{Kooyman1972,Falke1985,Hooker2005},
similarly to otariid seals, which are observed to exhale in the ascending phase of their dives \citep{Hooker2005}.
\added{An additional explanation to these behaviours could, however, be the intention of reducing dive locomotion cost.}
\deleted{The intention of reducing the locomotion cost of the dive could be however an additional explanation.
of the peculiar practice of phocid seals that are observed to exhale before diving.}
\deleted{Oxygen storage in phocid seals is in fact principally located in the blood and in the muscular tissues with only 5\% in the lungs.}

\begin{figure}
\begin{center}
\includegraphics[width=\columnwidth]{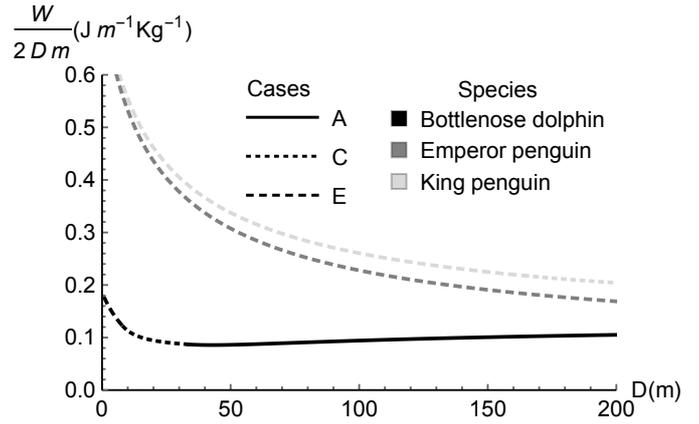} 
\caption{Locomotion cost per unit of covered distance and diver body mass as function of the dive depth for bottlenose dolphins\replaced{,}{ and} emperor \added{and king} penguins swimming at a constant speed of 1.8~m s$^{-1}$.
We consider a bottlenose dolphin with mass $m=177$~Kg, drag constant $C=4.15$~kg m$^{-1}$ and body volumes $R_g= 0.0493$ and $R_t =  0.9809$\added{, a king penguin with mass $m=11.4$~Kg, $C=0.50$~kg m$^{-1}$ and body volumes $R_g= 0.1322$ and $R_t =  1.0069$} and an emperor penguin with mass $m=30$~Kg, $C=0.60$~kg m$^{-1}$ and body volumes $R_g= 0.1369$ and $R_t =  1.0069$.}
\label{fig:WvsD}
\end{center}
\end{figure}

From Figs.~\ref{fig:work1} and \ref{fig:work3D} we can see that the measured dolphin's body values $R^\mathit{meas.}_t = 0.9809$ and $R^\mathit{meas.}_g = 0.0493$ considered here are close to the minimal region demonstrating dolphins' good adaption for dives around $100$~m deep.
This good adaption is not limited to these depths but also for a larger range value as we can see from the plot in Figure~\ref{fig:WvsD}. Here we plot the value of $W$ divided by diver mass and the distance covered (equal to $2 D$) as a function of the maximal depth.
We see that, except for dives close to the surface, the normalised locomotion cost is almost constant, demonstrating a good adaption of dolphin body characteristics for dives down to 200~m and more.
For comparison, we plot the normalized locomotion cost for emperor \added{and king} penguins\replaced{.}{, where} 
In particular we consider the example of a specific emperor penguin with $m=30$~Kg, $V^0_g=4$~l and $C=0.60$~kg m$^{-1}$ (data from \citet{Sato2010}, corresponding to the values $R_g= 0.1369$ and $R_t =  1.0069$),
\added{and a king penguin with $m=11.5$~Kg, $V^0_g=1.48$~l and $C=0.50$~kg m$^{-1}$ (data from \citet{Sato2002}, corresponding to the values $R_g= 0.1322$ and $R_t =  1.0069$).}
For all divers we consider the same cruise velocity of $1.8$~m s$^{-1}$.
As mentioned above, due to their low average tissue density compared to the salty water ($R_t>1$), emperor penguins are always positively buoyant at any depth and so fall into case E, with a glide during the entire ascent. 
\added{In the case of the king penguin, this is not completely true. If the dive is too deep, even if they are always positively buoyant, the animal has to actively swim in the first part of the ascent against the drag (case C). This is in agreement with the observation reported by \citet{Sato2002} (see next section for a quantitative comparison between our prediction and these data).}
As expected from Figure~\ref{fig:work1}, for penguin\added{s} the thrust work is greater, and decreases progressively with the depth with a global energy economy worse than dolphins.
\added{From the consideration above, this can be interpreted as consequence of the absence of neutral buoyancy during the dive. 
Dolphins have neutral buoyancy (without considering the drag) at depth $d \approx 15$~m with the presence of the prolonged glides in both descending and ascending phases.
In these phases, weight and buoyant forces are used best against the drag, resulting in a much lower  and the locomotion cost. 
This is not completely true for shallow dives, where the prolonged glide during the descent cannot be present and the dolphin glides for the entire ascent (case E) causing an increase in the dive cost.}

The optimal $R^\mathit{best}_t$ value minimising $W$ depends only marginally on the drag force (on $F_D^2/(m g)^2$ or higher terms).
This means that dolphin characteristics are not only optimal for a wide range of maximal depths but also for a wide range of swim velocities.

\subsection{Gliding region extension \label{sec:depths}}
For a defined set of parameters, Eqs.~(\ref{eq:deq}) can be used to predict the depths where the diver should begin their glide.
\added{We can perform a first quantitative comparison for king penguins, for which a large set of glide start depth values $d^\mathit{meas.}_A$ is reported \citet{Sato2002}. 
The depth of active swim cessation in the ascent depends on pre-dive lung filling and pitch angle assumed during the dive. 
Assuming the maximal air lung filling and vertical dives, Eqs.(\ref{eq:deq}) provide the maximum reachable value of $d_A$. 
Considering the king penguin characteristics and their typical swim velocity from \citet{Sato2002} ($m=11.5$~Kg, $V^0_g=1.48$~l and $C=0.50$~kg m$^{-1}$,  $v = 2$~m s$^{-1}$), we find that $d_A$ values should be less than 111~m.
This value is well in agreement with the observed $d^\mathit{meas.}_A$ which are all less than 100~m \citep{Sato2002}.

A much more quantitative comparison can be done with dolphins for which observations of pure vertical dives are available.}
For a 100~m deep dive by the same dolphin considered in the previous section, Skrovan \text{et. al} observed a glide during the descent starting at a depth of $d^\mathit{meas.}_{D} =67.5$~m preceded by an active swimming phase with a cruise velocity of 1.7~m s$^{-1}$, and a glide during the ascent starting at $d^\mathit{meas.}_{A} =5.5$~m preceded by an active swimming phase with a cruise velocity of 1.9~m s$^{-1}$.
For the same conditions, our model predicts the values $d_{D}=30.5$~m and  $d_{A}=8$~m, which are significantly different from the measured values.
A possible partial pre-dive filling of the lungs, and then a reduced value of $R_g$, causes a reduction in both critical depths, ruling this out as an explanation for the inconsistency with the observed values.
The most probable cause of this discrepancy is an incorrect evaluation of the drag force, i.e., the value of the drag constant $C$ or of the ratio $\lambda$ between the active and passive swimming drag force.
This can be easily understood from  Eqs.~(\ref{eq:deq}). 
An increasing of $F_d$ causes an increasing of $d_D$ and a decreasing of $d_A$.
Other parameters of Eq.~(\ref{eq:deq}) cannot be varied easily.

For our predictions, we use the drag constant for uninstrumented dolphins ($C=4.15$~kg~m$^{ -1}$), but the observations of the gliding regions are relative to instrumented dolphins for which  \citet{Skrovan1999} observe a drag increase of more than 4 times (corresponding to $C=16.9$~kg~m$^{ -1}$, extracted from the data points in \citet{Skrovan1999}).
With this drag constant, we should observe almost no gliding phases except for the very end of the ascent.
The same results can be obtained by taking a value of $\lambda$ several times larger than one.
The problem is, in fact, more complex because carrying instruments could affect the animal behaviour in an non-trivial way, as discussed by \citet{vanderHoop2014}.
In addition in our simple model we do not consider the stroke-and-glide swimming that could introduce an additional dependency on $v$ of the average values of $C$ and $\lambda$.
A deeper understanding of passive and active drag is required for a better prediction of glide phase extent.

We can, however, provide the maximal value that $C$ and $\lambda$ can assume from comparison between obverved values and the simple formulae of the critical depths (Eqs.~(\ref{eq:deq})).
Their single values cannot however be disentangled from each other. 
Independently from the possible approximations, we find that  $\lambda$ cannot be higher than 1.5 (i.e. $C$ cannot be 1.5 time larger than the value for uninstrumented dolphins), much less than the value considered in some past studies which indicates $\lambda = 3-7$ \citep{Lighthill1971,Webb1975,Fish1998,Skrovan1999,Anderson2001,Fish2014}.

\subsection{Swim velocity and optimisation \label{sec:resE}}
As announced in Sec.~\ref{sec:totalE}, the choice of the swim velocity is a crucial point for the total energy cost $E$ due to the basal metabolism, which contribution is proportional to the total diving time $T = 2 D/v$, and the drag, whose increases with the square of the swim velocity $v$.

In the general case we have a non-trivial relationship between $E$, given by Eq.~(\ref{eq:Etot}), and $v$.
This is not completely true in cases D, E and F, where a simpler expression for $E(D,v)$ can be found due to absence of any critical depth in the expression of $W$. 
For cases D and E the diver is always negatively or positively buoyant and the expression of $W$ is simple (Eqs.~(\ref{eq:w_caseDE})).
The minimisation of $E$ with respect to $v$ leads to the optimal cruise velocity
\begin{equation}
v^\mathit{best} = \sqrt[3]{\frac {\varepsilon B} C}'
\label{eq:v_caseDE}
\end{equation}
\added{which is independent of maximal dive depth.}
This expression was previously found by \citet{Sato2010} for the study of the optimal speed of emperor penguins, which are always positively buoyant
and are included in the case E where $F_b-F_w>F_f$.

For case F, the expression of the optimal speed is slightly different due to the thrust work required for both descent and ascent. 
\added{Considering Eq.~(\ref{eq:Etot}) and the expression of $W$ for case F, Eq.~(\ref{eq:w_caseF}), we obtain an optimal velocity solution very similar to Eq.~(\ref{eq:v_caseDE}) with}
the denominator term $C$ in Eq.~(\ref{eq:v_caseDE}) replaced by $2 C$.
N.B. this is also the optimal velocity for a horizontal displacement. 

No simple analytical formula of  $v^\mathit{best}$ can be obtained for the other cases and numerical methods (Mathematica software in our case \citep{Mathematica}) must be applied for solving the equation $\partial E / \partial v = 0$.
To better compare our results with previous studies,  instead of $E$ we consider the cost of transport per units of mass and covered distance COT $=E/(2 D\ m)$.
In addition to $m, R_t, R_g$ and $C$ from \citet{Skrovan1999}, 
the metabolic and propulsion efficiencies $\varepsilon_m=0.25$ and $\varepsilon_p=0.86$ \citep{Fish1999}
are used and where aerobic efficiency only is taken into account for $\varepsilon$. 
This is justified by post-dive blood lactate measurements in dolphins \citep{Williams1999} showing an increase in anaerobic process only after significantly long dives (longer than 200~s), not considered here. 

For the value of $B$, we consider the basal metabolic rate reported by \citet{Yazdi1999} ($B/m=2.15$~W Kg$^{-1}$) measured for trained dolphins in a pool but considering some precautions.
During the dives, dolphins, 
like other breath-hold divers, are subject in fact to a physiological response to diving (diving reflex) with bradycardia, reduction and redistribution of the blood flow, etc.
Specifically, a reduction in heart rate to 63.4\% is observed  in dolphins \citep{Williams1999} with an expected substantial reduction of $B$.
To roughly take into account the diving reflex, we also examine the expected value of COT corresponding to half of the basal metabolic rate value reported in literature ($B/m=1.07$~W/Kg), which is also approximately equal to basal metabolism predicted by the Kleiber's law \citep{Kleiber1947,Yazdi1999}.

\begin{figure}
\begin{center}
\includegraphics[width=\columnwidth]{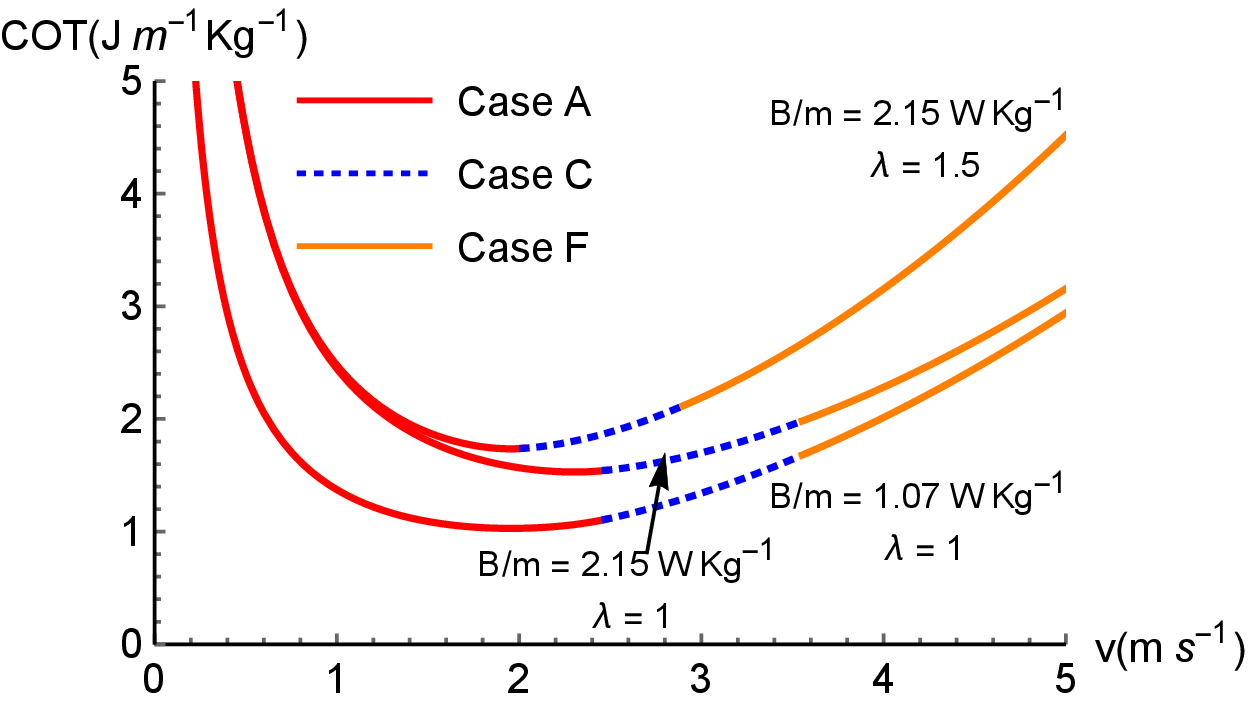} 
\includegraphics[width=\columnwidth]{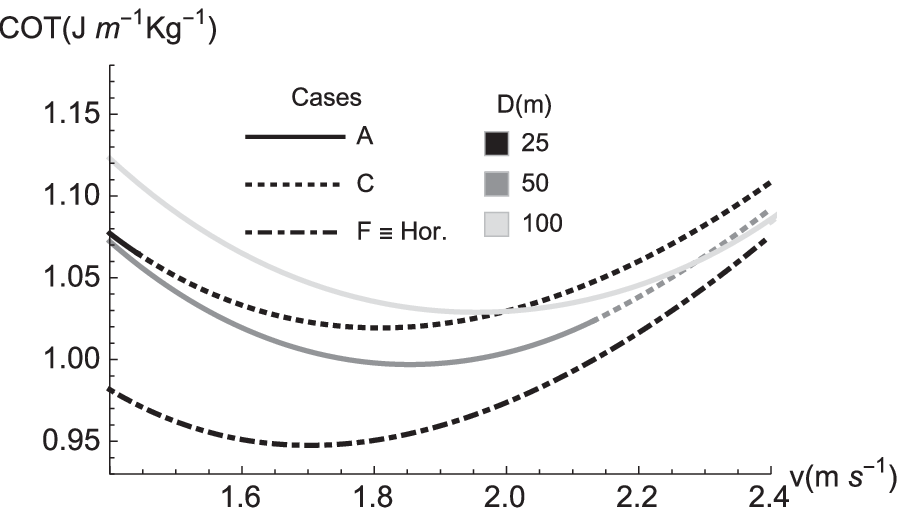} 
\caption{ Top: Cost of transport as a function of dive velocity $v$ for a bottlenose dolphin of mass $m=177$~Kg, drag constant $C=4.15$~kg m$^{-1}$ and body volumes $R_g= 0.0493$ and $R_t =  0.9809$ diving to a maximum depth of $D=100$~m. 
Different curves correspond to different assumptions for the basal metabolic rate $B$ and the active to passive drag ratio $\lambda$.
The presence of critical depths depends on the value of $v$ (both critical depths present: case A; only one present in the ascent: case C; none present: case F). 
Bottom: Close-up of the minimum region where the dependency on $D$ is more visible and where we assumed $B/m=1.07$~W kg$^{-1}$ and $\lambda = 1$.
}
\label{fig:COTdolphin}
\end{center}
\end{figure}

In Figure~\ref{fig:COTdolphin} we plot the COT for the same dolphin individual considered in the previous section for a 100~m depth dive for different combinations of  values of $B/m$ ($B/m=2.15$ and 1.07~W Kg$^{-1}$) and $\lambda$ (1 and 1.5).
Changes in dive velocity determine the existence or not of critical depths and so which dive case applies.  
Here in particular, cases A, C and F must be considered. 
For $B/m=2.15$~W Kg$^{-1}$ and $\lambda=1$ the optimal velocity, solution of the equation  $\partial E/ \partial v=0$, is  $v^\mathit{best} =2.3$~m s$^{-1}$. 
This value is quite different from the value of about 1.8~m s$^{-1} $reported by \citet{Skrovan1999}.
When the basal metabolic rate value of $B/m=1.07$~W Kg$^{-1}$ is considered to include the diving response, an optimal velocity of $v^\mathit{best} =1.95$~m s$^{-1}$ is obtained, very close to the measured values.
An underestimation of the drag force can also produce an artificial decreasing of $v^\mathit{best}$.
We remember that we consider here a value of the drag constant for uninstrumented dolphins when the observations are made on instrumented dolphins.
If we take the value of $C$ for instrumented dolphins from \citep{Skrovan1999} ($C=16.9$~kg m$^{-1}$), we get the very low value of $v^\mathit{best} =1.3$~m s$^{-1}$ -- very different from the observed value, as we found for critical depths (Section~\ref{sec:depths}).
Similarly, if we consider larger values of $\lambda$, a smaller $v^\mathit{best}$ is obtained.
In fact, when the $\lambda$ upper limit of 1.5 is adopted (obtained from the observation on the critical depths in Sec.~\ref{sec:depths}), the value $v^\mathit{best} =1.95$~m s$^{-1}$ is found (see Fig.~\ref{fig:COTdolphin}), the same value obtained using half the $B$ value from the literature.

The cause of the disagreement between observed and predicted swim velocity when the literature value of $B$ is used is probably due to a combination of the effect of the diving response and the underestimation of drag.
However, considering i) the recent studies that indicate a swim speed reduction of only 11\% for instrumented dolphins \citep{vanderHoop2014}, (carrying different apparatus than in \citet{Williams1999,Skrovan1999}),
 ii) the evaluation of $\lambda < 1$ reported in some recent articles \citep{Hind1997,Barrett1999,Borazjani2008} and iii) the observed heart rate reduction of diving dolphins \citep{Williams1999}, we suggest that the diving reflex plays the major role and a reduced value of $B$ should be considered in favour of an increase of $\lambda$ or $C$ values.
For simplicity, the following calculations use the values $B/m=1.07$~W kg$^{-1}$, $\lambda = 1$ and  $C=4.15$~kg~m$^{ -1}$ (from uninstrumented dolphins).

\begin{figure}
\begin{center}
\includegraphics[width=\columnwidth]{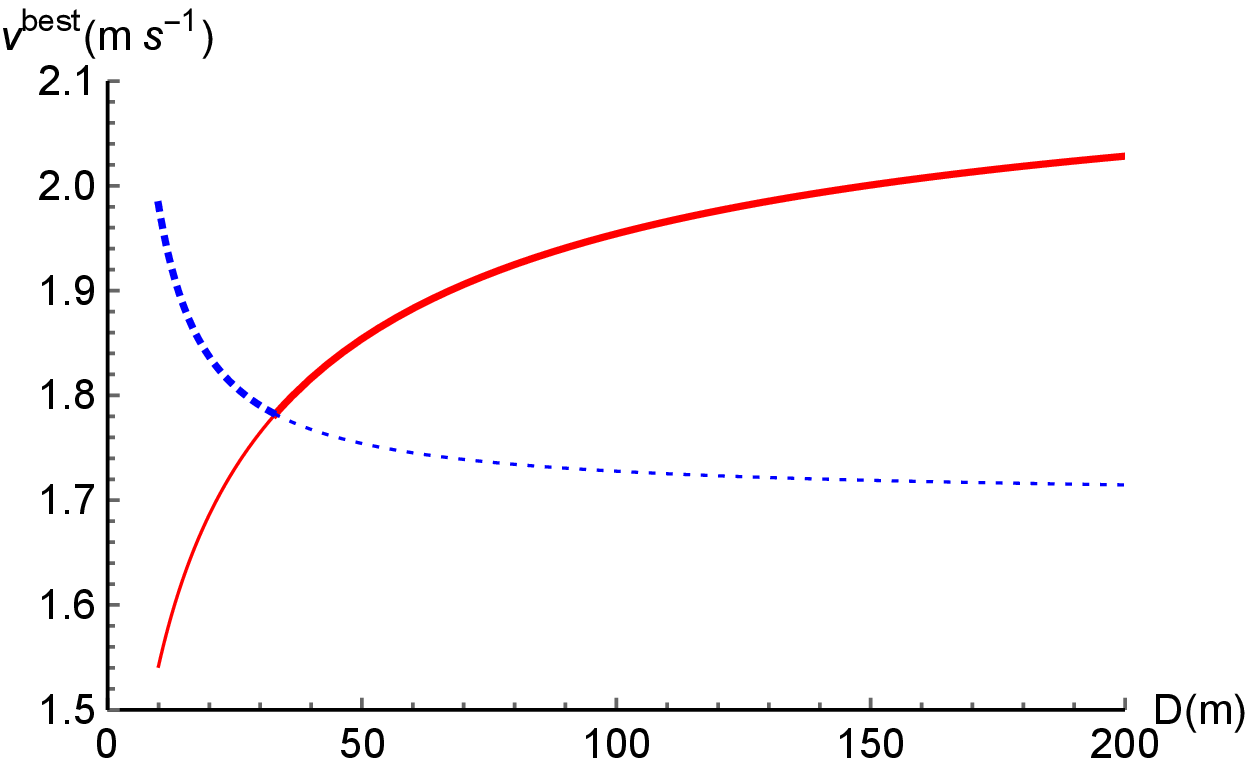} 
\includegraphics[width=\columnwidth]{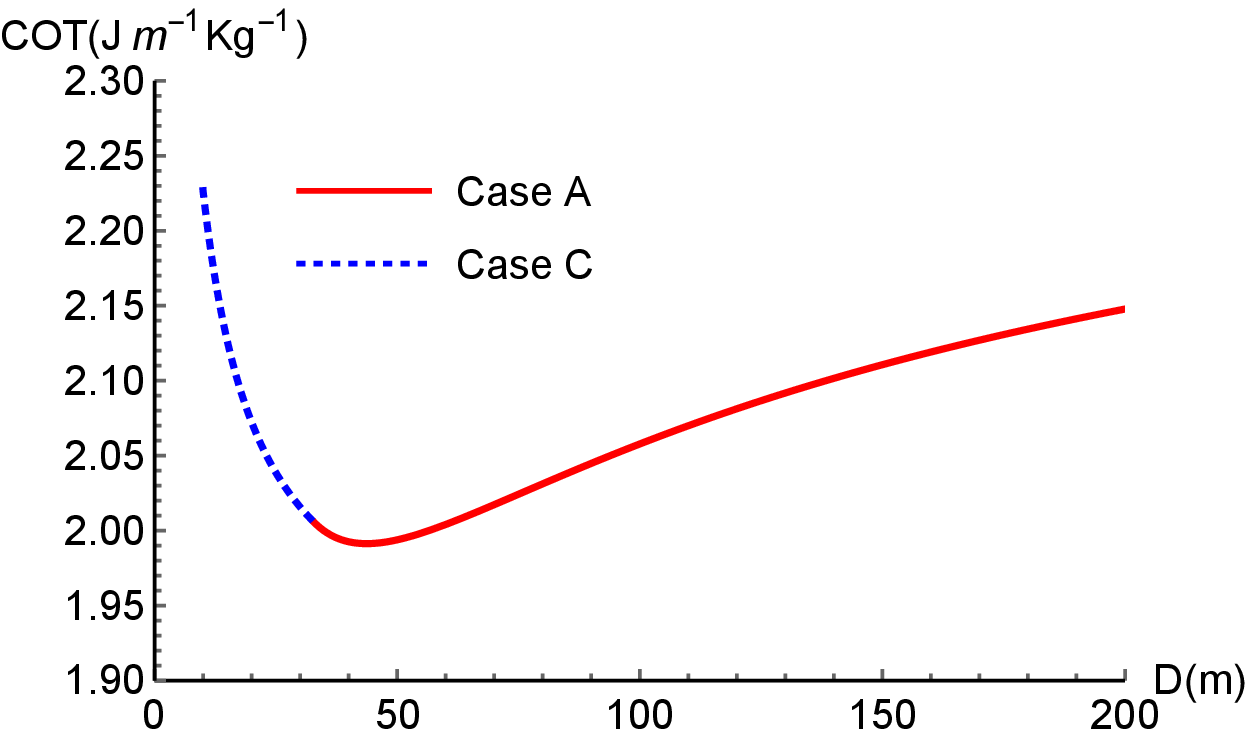} 
\caption{Optimal velocity dependency (top) and correspondent COT (bottom) on the maximal dive depth for a bottlenose dolphin of mass $m=177$~Kg, drag constant $C=4.15$~kg m$^{-1}$ and body volumes $R_g= 0.0493$ and $R_t =  0.9809$. Solution case A or C has to be considered. Both cases are represented and the pertinent values of $v^\mathit{best}$, which depends on $D$, are indicated by thick lines.} 
\label{fig:VvsD_dolphin}
\end{center}
\end{figure}

The dependency of the COT on maximal dive depth is marginal, as it can be observed in Figure~\ref{fig:COTdolphin} (lower panel) where we plot the COT for dives with $D=25, 50, 100$~m.
Optimal dive velocity dependency on maximal dive depth and the corresponding value of COT are presented in Figure~\ref{fig:VvsD_dolphin}.
Different depths correspond to different cases.
As expected, large values of $D$ fall into case A. For \added{the particular dolphin individual considered,} dives of less than 33~m fall into case C (no prolonged glide in the descent).
Presence or absence of prolonged glides and dependency on their extent produce a non-trivial dependency of the optimal speed on the maximal depth.
$v^\mathit{best}(D)$ varies about 10\%, in the 1.8 and 2.0~m s$^{-1}$ range, similar to the observed values \citep{Skrovan1999}.
Differently to $W$ (Fig.~\ref{fig:WvsD}), the cost of transportation calculated for $v= v^\mathit{best}$ has a minimum with respect to the maximal dive depth $D$, which in this case is around $D \approx  40$~m.

We can study the dependency of $v^\mathit{best}$ and the corresponding COT on changes of $R_t$ and $R_g$ ratios, as we did for locomotion work.
In this case, however,  we have a more complex problem. 
$v^\mathit{best}$ is the solution of $\partial E/ \partial v=0$, where $E$ must correspond to the pertinent case (A to F), whose selection depends on the $F_d$ value and then on $v$.

\begin{figure}[ht]
\begin{center}
\includegraphics[width=\columnwidth]{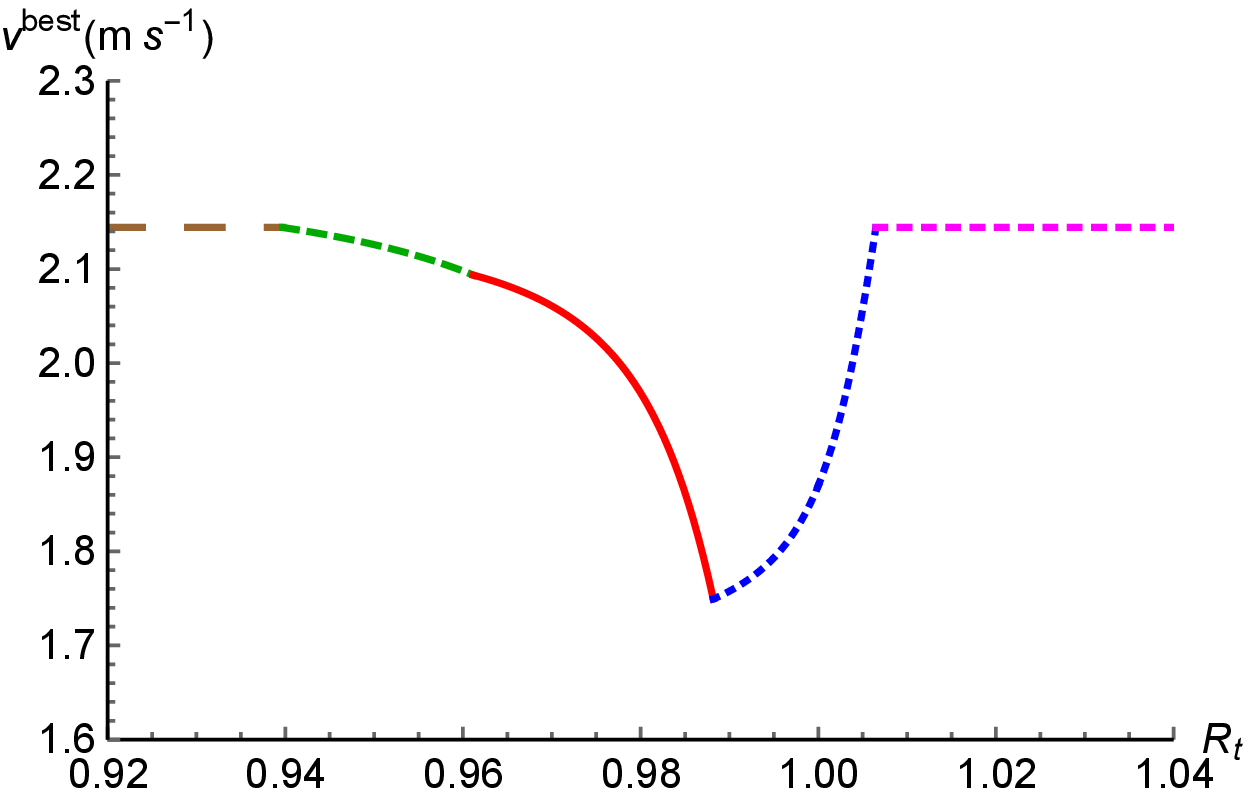} 
\includegraphics[width=\columnwidth]{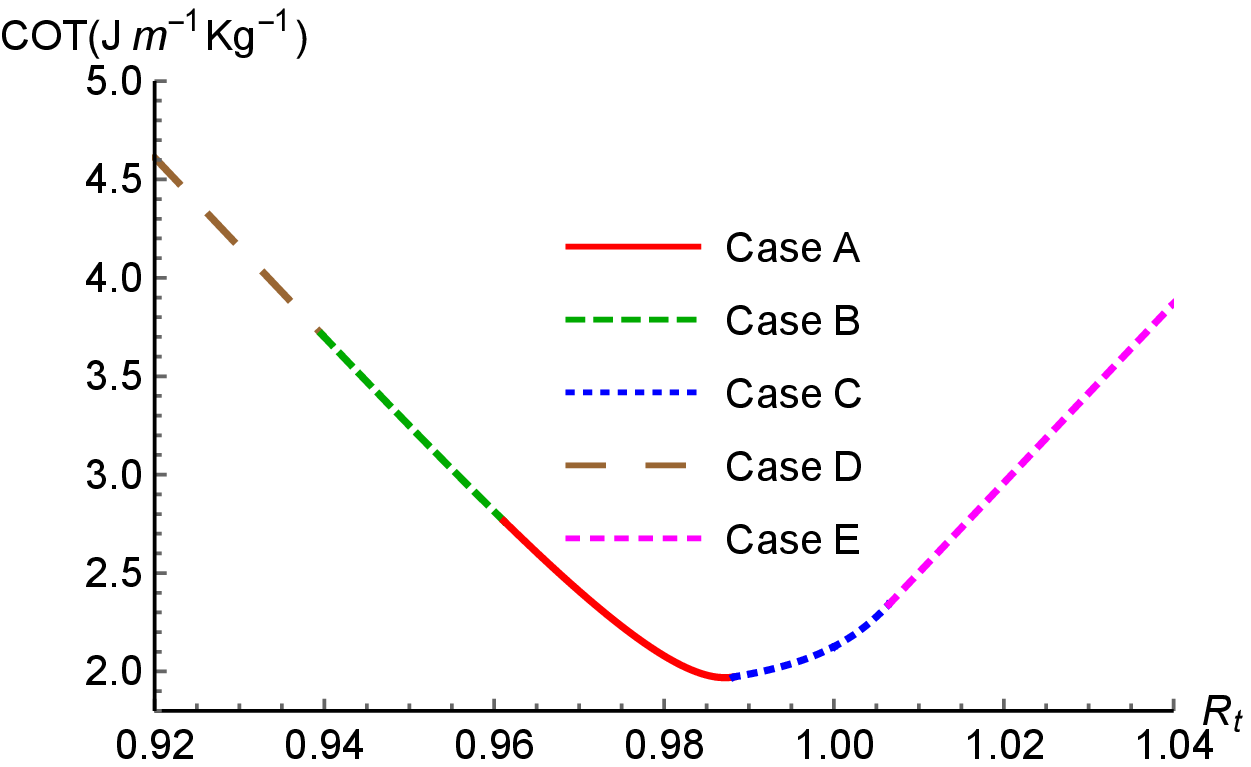} 
\caption{Optimal velocity dependency (top) and correspondent COT (bottom) on the ratio $R_t$ for a bottlenose dolphin of mass $m=177$~Kg and drag constant $C=4.15$~kg m$^{-1}$ diving to a maximum depth of $D=100$~m keeping $R_g$ constant at 0.0493 ($=R^\mathit{meas.}_g$).}\label{fig:VvsRi}
\end{center}
\end{figure}

As we can see in Figure~\ref{fig:VvsRi} (top panel), when we vary $R_t$ keeping $R_g = R^\mathit{meas.}_g$ constant, $v^\mathit{best}$ varies up to 15\% in the region $R_t \sim 1$ (when prolonged glides are present).
When $R_t$ is relatively far from one, the diver is extremely positively or negatively buoyant (cases D and E) and the optimal velocity expression is given by the Eq.~(\ref{eq:v_caseDE}), independent of the magnitude of $R_t$ and $R_g$ values.

\begin{figure}
\begin{center}
\includegraphics[width=\columnwidth]{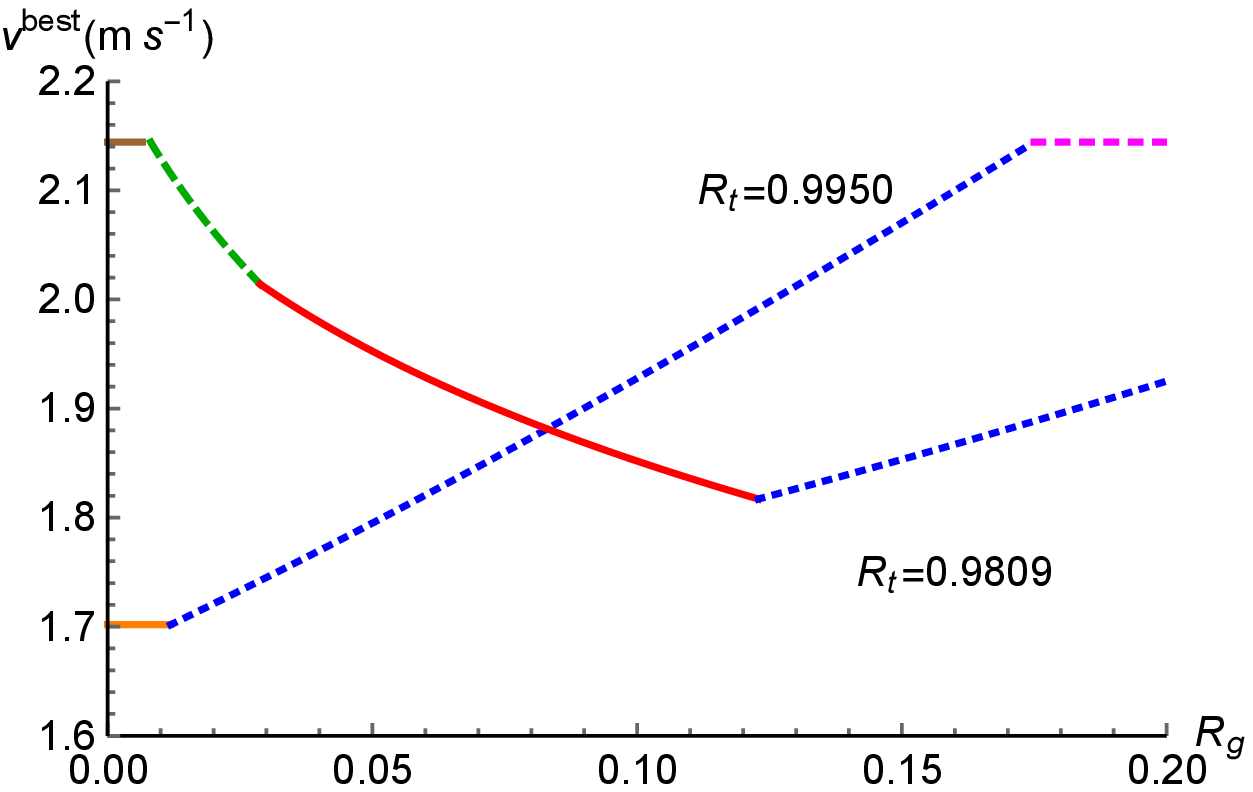} 
\includegraphics[width=\columnwidth]{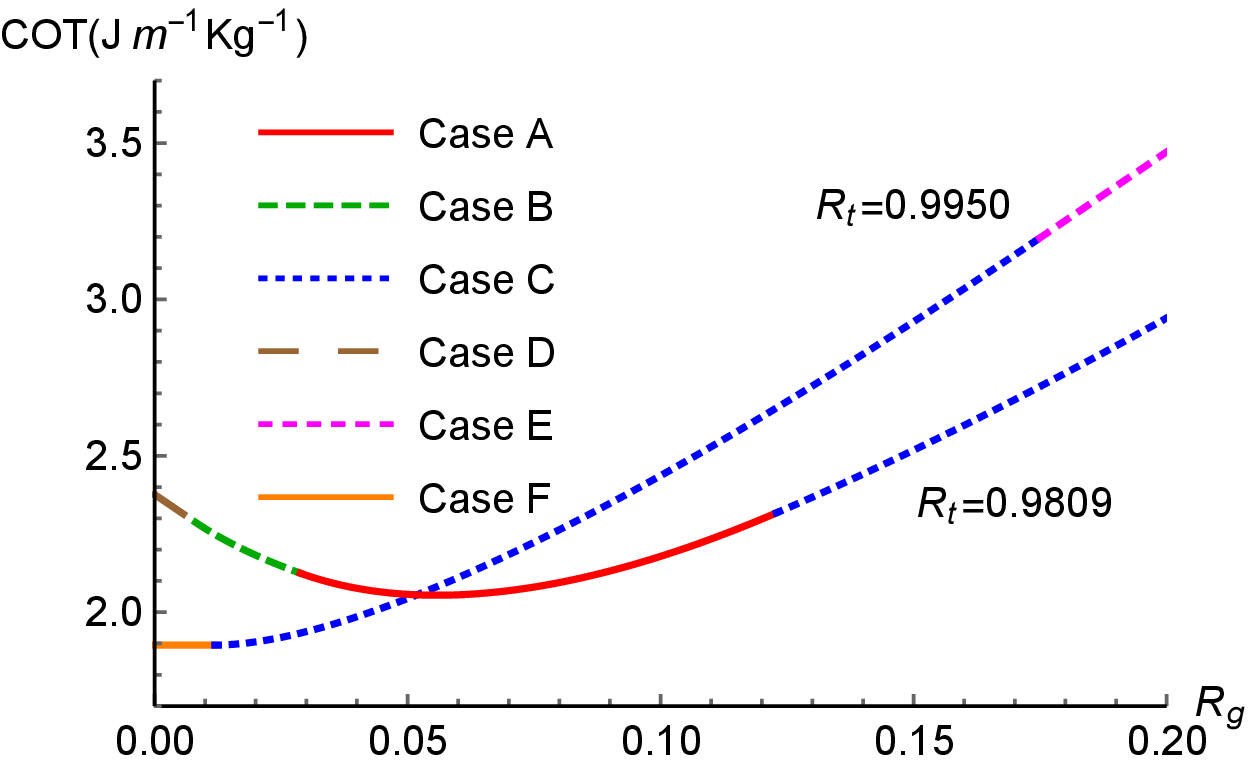} 
\caption{Optimal velocity dependency (top) and correspondent COT (bottom) on the ratio $R_g$ for a bottlenose dolphin of mass $m=177$~Kg and drag constant $C=4.15$~kg m$^{-1}$ diving to a maximum depth of $D=100$~m keeping $R_t$ constant at 0.9809 ($=R^\mathit{meas.}_t$) and 0.9950.}\label{fig:VvsRc}
\end{center}
\end{figure}
 
\begin{figure}
\begin{center}
\includegraphics[width=\columnwidth]{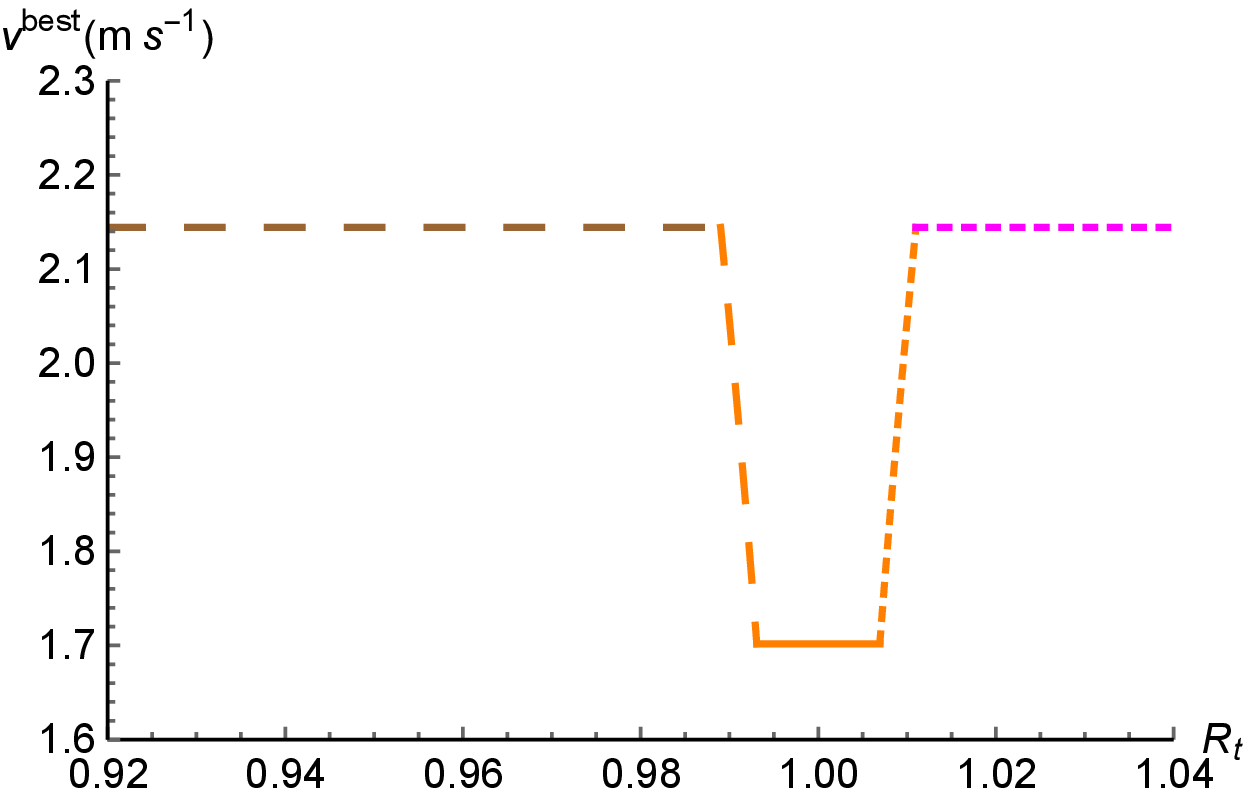} 
\includegraphics[width=\columnwidth]{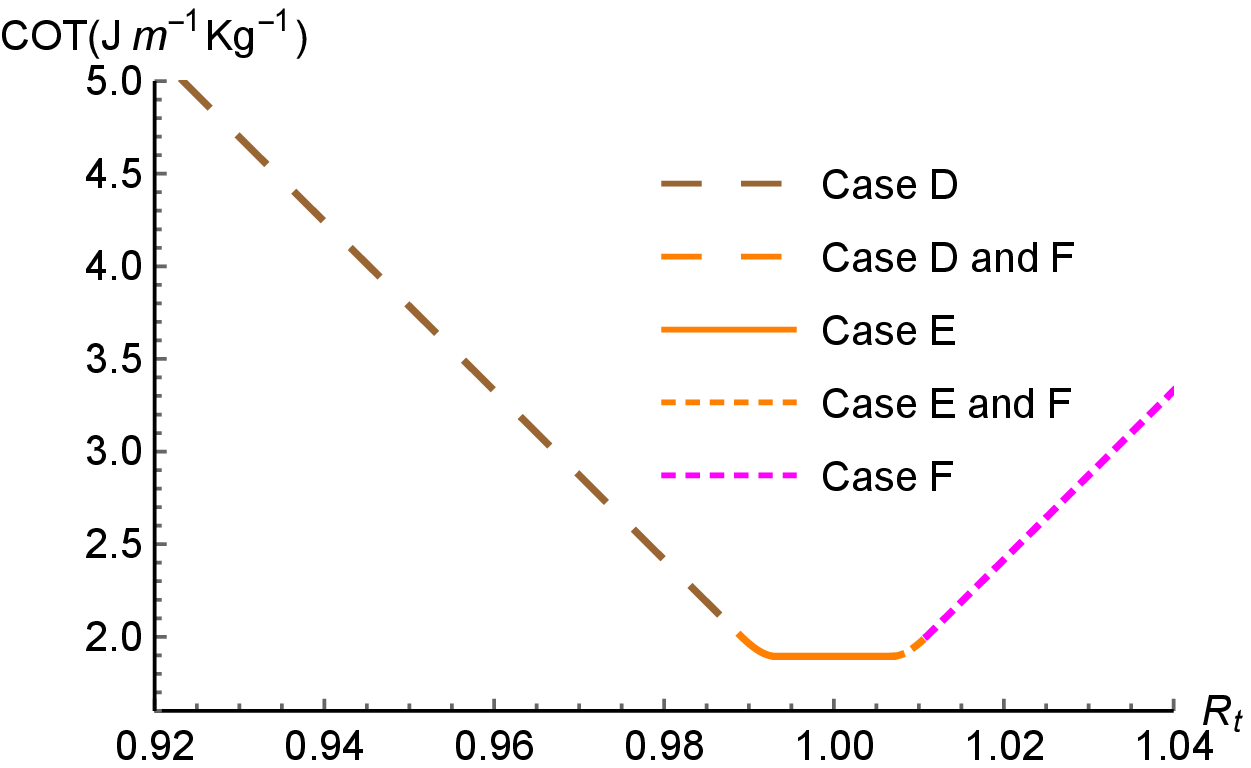} 
\caption{Optimal velocity dependency (top) and correspondent COT (bottom) on the ratio $R_t$ for a bottlenose dolphin diving to a maximum depth of $D=100$~m for the extreme case $R_g = 0$.}\label{fig:VvsRi_Rc0}
\end{center}
\end{figure}

As we can see in Figure~\ref{fig:VvsRc} (top panel), variation in $R_g$ keeping $R_t$ constant causes the switch from one case to another with extreme cases E and D or F, depending on the considered value of $R_t$. 
\added{Close to neutral buoyancy (}$R_t$ values enough close to one\added{)}, case F in fact replaces case D.
This is visible also in Figure~\ref{fig:VvsRi_Rc0} (top panel) where the variation of $v^\mathit{best}$ with respect on $R_t$ is presented for the extreme case $R_g = 0$.

The optimal cost of transportation corresponding to $v= v^\mathit{best}$ for different values of $R_t$ and $R_g$ is presented in the lower panels of Figures~\ref{fig:VvsRi} -- \ref{fig:VvsRi_Rc0}. 
Dependency of the COT on ratios $R_t$ and $R_g$ is similar to $W$, with a global minimum close to the region $R_g = 0, R_t = 1,$ \added{i.e. for almost neutrally buoyant divers with a very small gas-filled body volume}.
This confirms the advantage of exhaling before diving dive, observed in phocid seals, whose $R_t = \rho/ \rho_t$ value can be close to one \citep{Aoki2011,Sato2013}. 
\added{We remember that unlike other marine mammals, air stored in phocid seal lungs represents only the 5\% of the total stored oxygen, which is mainly located in the blood and in muscular tissue. In this case, air exhalation does not reduce too much the total available oxygen but does allow for reducing the COT.}

As for $W$, the minimal region corresponds to the case F, for which the expression of COT becomes very simple and does not depend on physical characteristics of the diver: 
\begin{equation}
COT^F(v = v^\mathit{best}) = \frac 3 m \sqrt[3]{\frac{C B^2} {4 \varepsilon}}. 
\end{equation}

For the extreme condition where $R_g = 0$, unlike $W$, in addition to the possible cases D, E and F, we have two transition regions corresponding to case F limits with cases D or E, where $d_A = 0$ or $d_D = D$, respectively. 
Unlike $W$, we are in fact adjusting $v$ to minimise the COT for each value of $R_t$ which influences case selection.

The presence of the COT minimum close to $R_t=1$ when $R_g=0$ confirms the results from \citet{Miller2012,Sato2013}, where, when $R_g=0$ \added{(equivalent to considering buoyancy not dependent on $d$)}, the resulting locomotion cost is minimal at neutral buoyancy \added{but where no variation of the buoyant force is considered}.
Here we extend this result to a larger range of values of $R_t$ corresponding to the region defined by Eq.~(\ref{eq:Ri_Rc0}) with
\begin{equation}
F_d = C (v^\mathit{best})^2 = \sqrt[3]{\frac{C \varepsilon^2 B^2} {4}}.
\end{equation}

\subsection{Mass dependency of optimal dive velocity}

\begin{figure}
\begin{center}
\includegraphics[width=\columnwidth]{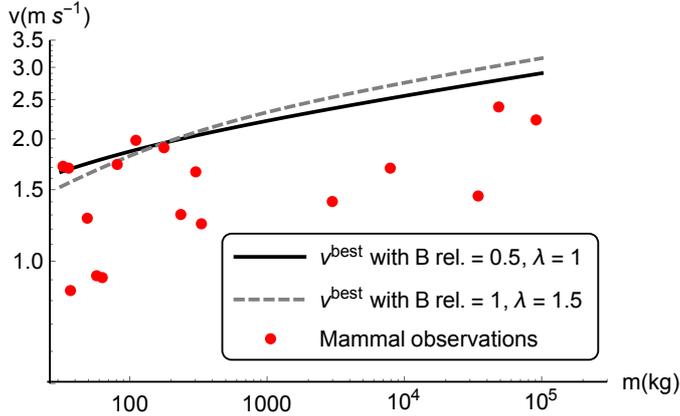} 
\caption{Breath-hold diver velocities for a dive 100~m depth as function of diver mass. Theoretical predictions with different  assumptions are compared to available observations in mammals \citep{Watanabe2011}.} 
\label{fig:VvsM}
\end{center}
\end{figure}

A more general comparison between the prediction of our model and observed values can be done by considering different diver masses. Like  \citet{Watanabe2011}, we introduce a mass dependency on the different parameters of Eqs.~(\ref{eq:W_gen}--\ref{eq:Etot}).
For the drag constant $C$ we consider a variation of $m^{3/5}$ from the dependency of the diver surface ($m^{2/3}$) and the drag coefficient ($m^{-1/15}$).
For the basal metabolic rate we take the classic dependency of $m^{3/4}$ from the Kleiber's law, keeping as reference half the value of $B$ ($B$ relative = 1/2) of resting dolphins at the surface \citep{Yazdi1999} to take into account the physiological response to diving.
From observations \citep{Kooyman1998} and basic principles, it is difficult to make an estimation of lung volume variation with the mass. 
We consider then, no dependency on $m$ for $R_g$  as for the efficiency $\varepsilon$ and we retain the value $\lambda = 1$.

In Figure~\ref{fig:VvsM} we plot $v^\mathit{best}$ for a dive with $D=100$~ m as function of $m$ for a large range of diver masses, all corresponding to case A where both critical depths are present. 
As we can see, contrary to \citet{Watanabe2011}, no exponential law can be extrapolated. 
Comparing our predictions to the observed velocities of mammals \citep{Watanabe2011}  (which generally glide in both ascending and descending phases)
we see that our predictions are in quite good agreement for small values of $m$ but are systematically higher for larger masses.
In their study \citet{Watanabe2011} suggest a possible reason for the observed large difference in seabird and mammal swim velocities (both endotherms) could be due to the large difference in thermoregulatory costs between the species. Similarly, a reduction in thermic loss coupled to a strong diving reflex could explain large mammals' small swim velocities. 
The discrepancy between predicted and observed velocities could also come from the approximations made in our model, such as the assumptions made for $\lambda$ and $B$.
When a greater value of $\lambda$ ($\lambda  = 1.5$ instead of 1) and any reduction of $B$ due to the diving reflex ($B$ relative = 1) are considered, the  disagreement between predictions and observations increases. 
This suggests that the origin of this discrepancy is elsewhere, such as i) the value of $B$ discussed above, or ii) neglect of possible accelerations during prolonged glides or iii) rough approximations of the dependency of the different parameters, in particular $C$ and $R_g$, on diver mass.
Further studies are required to clarify this point.

\section{Conclusions}
We presented a theoretical model for breath-hold animal dives. 
\added{In accordance to previous studies \citep{Wilson1992,Skrovan1999,Hansen2004,Sato2010,Aoki2011,Watanabe2011,Miller2012}}, starting from basic principles of forces acting on the diver, we calculated the work of the thrust to evaluate energetic locomotion dive cost.
Unlikely to previous studies, \deleted{considering simple cases with a constant cruise velocity during the dive and few approximations}, we were able to take into account \added{accurately for the first time} the presence of prolonged glides during the ascent and descent, calculate their extent and study their impact on the total locomotion cost.
In particular we analyse locomotion cost dependency on different dive parameters, namely diver body characteristics (mass, lung volume, buoyancy, etc.) and dive characteristics (cruise velocity and maximum depth).
To calculate total dive energy cost, we also include the basal metabolism and efficiency for transforming chemical energy to propulsion. We studied total dive cost dependency on cruise velocity and the choice of optimal swim speed to minimise dive cost as function of the different parameters.

\added{We demonstrate that both locomotion and total energy cost are minimised for divers which pass through a condition of neutral buoyancy during the dive, generally implying the presence of prolonged gliding phases in both ascent and descent.
This is in agreement with past results  \citep{Miller2012,Sato2013} where, when the buoyant force is considered constant during the dive, energy cost is minimised for neutral buoyancy.
In particular, our model confirms good physical adaption of dolphins for dives of $10-200$~m deep due to their small body tissue density over mass ratio,} \deleted{We found that the body characteristics of bottlenose dolphins, in particular the ratio between the body tissue volume and their mass, are well optimised for dives of $50-200$~m,} contrary to the case of emperor \added{king} penguins.
\added{The presence of prolonged glides implies a non-trivial dependency of optimal speed on maximal dive depth, and extends previous findings \citep{Sato2010,Watanabe2011} who found no dependency on dive depth.}
\deleted{From the dependency of the locomotion cost on the lung volume, we also could provide an additional explanation on the observed exhalation of air of seals before diving, normally associated to the reduction of decompressions sickness risks.}
\added{Locomotion and total energy cost of the dive are further diminished by reducing gas-filled body parts volume for divers close to neutral buoyancy.}
In particular, this provides an additional possible explanation of the pre-diving exhalation of air observed in phocid seals to minimise dive energy cost, rather than the current explanation from the literature of decompression sickness risk reduction.

\deleted{We demonstrate that both locomotion and total energy cost are minimised when the gliding region are reduced.
In particular, we find that their global minima corresponds to divers close to neutral buoyancy with a volume of body parts filled with gas (lungs, oral cavities, etc.) close to zero.
This confirms in particular the results of the recent works of Miller 2012  and Sato 2013
When the maximal depth of the dive is varied, we find that the optimal speed has a non-trivial dependency on  $D$ due to combination of the effect of the basal metabolic rate and the presence and extension of the prolonged glides.}

\added{We also successfully compare the extent of gliding phases with observations of penguin and dolphin dives. In particular, from the quantitative comparison between the data for vertical dives }
\deleted{From the comparison between the observation of the extension of the region} of a bottlenose dolphin  \citep{Skrovan1999} and our model predictions, we estimate an upper limit of the ratio of active to passive drag of 1.5, confirming the findings of studies with robotic fish and simulations \citep{Barrett1999,Borazjani2008}.

The comparison between our findings and observed swim velocities of breath-hold mammals with different masses suggests that the physiological response to diving could play an important role in the choice of cruise velocity but it also shows the limitations of our model and its approximations.
In particular, the assumption of a constant cruise velocity during the dive and the dependency of the drag on the swim speed, including a better involvement of the stroke-and-glide swimming mode, and on other parameters, should be revisited and investigated.

\section*{Acknowledgments}
We would like to thank the team of the ``7$^\text{\`eme}$ Apn\'ee'' freediving club for stimulating the beginning of this study, and the group ASUR of the Institute of Nanoscience of Paris and many other persons for supporting its development.

\section*{References}


\bibliographystyle{model2-names}
\bibliography{breath-hold_divers_revised2}

\listofchanges

\end{document}